# Light scattering and surface plasmons on small spherical particles

Xiaofeng Fan[a], Weitao Zheng[a] and David J. Singh[a,b]

a. College of Materials Science and Engineering, Jilin University, Changchun 130012, China
b. Materials Science and Technology Division, Oak Ridge National Laboratory, Oak Ridge, Tennessee 37831-6056, USA

**Abstract**

Light scattering by small particles has a long and interesting history in physics. Nonetheless, it continues to surprise with new insights and applications. This includes new discoveries, such as novel plasmonic effects, as well as exciting theoretical and experimental developments such as optical trapping, anomalous light scattering, optical tweezers, nano-spasers, and novel aspects and realizations of Fano resonances. These have led to important new applications, including several ones in the biomedical area and in sensing techniques at the single-molecule level. There are additionally many potential future applications in optical devices and solar energy technologies. Here we review the fundamental aspects of light scattering by small spherical particles, emphasizing the phenomenological treatments and new developments in this field.

**Introduction**

The production, control, manipulation and use of light are at the core of many technologies. Light scattering plays key roles in all of these. Of course, the scattering of light by small particles has a long history, where it was studied in contexts such as cumulus clouds, the color of the sky and rainbows, and used in various glass artifacts and windows from the middle ages[1]. The remarkable fact is that such a classical topic is the basis of many fundamentally new and unexpected scientific and technological advances. The key is the current focus on the nanoscale and especially near-field effects at the nanoscale, while much of the older classical study was oriented towards the accessible far-field behavior.

More specifically, there have been fascinating developments in regard to the light scattering by nanosized particles, including metal particles and surfaces, where localized surface plasmons can be excited leading to optical resonance phenomena[2-5]. Small particles with surface plasmons can be used to detect the fluorescence of single molecules[6, 7], enhance Raman scattering[8], resonantly transfer energy of excitons[9], and create nanosized quantum amplifiers of optical energy. Potential practical uses include[10] small-scale sensing techniques[11, 12], numerous biomedical applications[13], manipulation of light for solar energy technologies[14], and others.

Here we provide a short review emphasizing the nano-optics of small particles, near field effects, and the fundamental theoretical basis for their description. We begin with a review of the classical light scattering theory for spherical particles based on the quasi-static (Rayleigh) approximation and the general Mie theory. Scattering by dielectric particles is discussed along with the new topic of optical trapping. We discuss plasmon resonances and light scattering on small metallic particles, which is a subject that has been renewed by a series of new findings, including anomalous scattering with an inverted hierarchy of resonances and Fano resonances. The breakdown of the general Drude model for dielectric function at very small particle sizes and the resulting effects are discussed. Finally, we review the stimulated radiation from the surface plasmons of small particles along with concepts for new kinds of lasers based on nano-lasing related to surface plasmons coupled to an active medium (so-called spasers).

We start with a summary of the basic concepts that remain useful in understanding light scattering, focusing on the case of spherical particles. Light scattering by small particles is one of fundamental problems of electrodynamics. As mentioned, it is a classical subject for which theory was developed long ago. This theory includes both the near and far field description. However, until recently the near field was inaccessible to experiments, and the interest was focused on far field effects. Now with advances in nano-technology and nano-optics, the richness of the near field theory is being exploited. This includes the production by scattering of very high light intensities with spatial variations shorter than the wavelength – a phenomenon that enables rich new physics, both linear and non-linear, at the nanoscale.

The physical understanding of light scattering by small particles began with the electric dipole concept, introduced by Lord Rayleigh in 1871[15]. One starts with the assumption that the electromagnetic phase is constant over the region of interest, which is natural since the size of small particle considered is less than the wavelength of light. Then the homogeneous field of the incident light induces a polarization, which in turn results in light scattering. Higher order scattering modes, such as quadrupole and octupole, are not considered at this level. The polarization (i.e. the induced dipole) of materials in response to electromagnetic fields is determined by the dielectric function.

The dielectric function of a material (at energies above the phonon energies) is determined by its electronic structure. It is practical to calculate dielectric functions from first principles band theory and often good agreement with experiment is found[16, 17]. However, for analyzing optical properties, it is often useful to approximate the optical properties of solids using the classical harmonic oscillator formalism introduced by Lorentz. In the Lorentz model, the dielectric function of non-conducting materials can be expressed as[1] $\varepsilon = 1 + \frac{f}{\omega_0^2 - \omega^2 - i\gamma\omega}$, where $f$ and $\omega_0$ are a phenomenological oscillator strength and frequency representing the bound electrons and $\gamma$ is a damping constant. This naturally leads to the Sellmeier formula for the refractive index, $1/(n^2-1)=-A/\lambda^2+B$, where $n$ is the refractive index, $\lambda$ is the wavelength and $A$ and $B$ are material dependent quantities. This formula and generalizations (e.g. to two or more oscillators) are very effective in fitting the optical constants of real materials[18].

Many physical phenomena can be very simply understood even in the simplest one oscillator theory. For example, the dispersion of light by prisms or water drops is explained by the frequency dependence of the refractive index. This follows the normal dispersive behavior (refractive index increases with energy) for materials like glass and water. This originates in the fact that the energy ($\omega_0$) of the effective oscillator for transparent materials such as these is generally much larger than the frequency of visible light[19]. For metals, the contributions of free electrons need to be added. This yields a model known as the Lorentz-Drude model. This model has[1] $\varepsilon = 1 - \frac{\omega_p^2}{\omega^2 + i\gamma_e \omega} + \sum_j \frac{f_j}{\omega_j^2 - \omega^2 - i\gamma_j \omega}$, where the sum on $j$ is over different oscillators. The free electron part is due to the electron plasma of the metal, which is described by the parameters $\omega_p$ and $\gamma_e$ which represent the resonant frequency and damping constant of bulk plasma.

It has long been recognized that all linear optical phenomena can in principle be modeled by solving the Maxwell's equations with known dielectric functions of the media. This understanding, while correct, by itself yielded relatively little direct physical insight into light scattering, at least prior to the development of modern computers and electromagnetic codes. This is because the vector electromagnetic equations resisted analytic solution, especially for complicated, but interesting, boundary conditions.

The earlier work on light scattering by small particles is mainly from Lorenz, Thomson and Clebsh[20]. Actually, the exact solution has been obtained by Clebsh in 1861 in his paper "Concerning reflection on a spherical surface" published in 1863, a year before Maxwell's work about electromagnetic theory of light. The breakthrough in understanding light scattering by spherical structures came from the work of Mie in 1908[21]. He obtained a general rigorous solution, on basis of the electromagnetic theory, for the optical scattering by a homogeneous sphere with arbitrary size in a homogeneous medium, whatever the composition of the sphere and medium. The Mie theory solution is also applied directly to the scattering by any number of spheres if the distance between particles is large enough so that there are no coherent phase relations among the scattered light from different particles[19].

The general Mie theory of optical scattering is very useful in practice. Initially, interesting problems, such as the origin of rainbows and the solar corona, could be directly answered on basis of the Mie solution[19]. So-called corrected Mie theory gives the light scattering by structures with other regular shapes, such as ellipsoids with any size and cylinders with arbitrary radius[19]. The basic optical properties of small particles made of different materials have been analyzed in detail via modifications of Mie theory, such as the Gans modification (for spheroidal particles, e.g. plasmonic gold and silver nanoparticles with elongated shapes) and the Maxwell-Garnett equations (providing an effective medium approach)[22]. Of course, there are also interesting questions related to light scattering by particles that cannot be directly described with Mie theory. An example is the electromagnetic hot spot between two nearby particles, which depends on coherency.

As mentioned, it is remarkable that even though more than 100 years have passed since the introduction of Mie's general theory of light scattering by a sphere, new and exciting physics associated with light scattering by small particles continues to be found[23-26]. Examples include the finding of giant optical resonances with an inverted hierarchy (e.g. the quadrupole resonance is more intense than the dipole) in scattering by small particles with negative dielectric susceptibility and weak dissipation[23], and anomalous scattering with the complicated near-field structures, such as the vortices, unusual frequency dependence, etc.[27, 28]. In addition, the Fano resonances, which are well known in quantum physics, were discovered in optics of small metallic particles[29-31]. Turning to nanotechnology and nano-optics, there is an increasing focus on control of optical energy in sub-wavelength structures – a field that is leading to many new ideas and remarkable experiment results[3, 32-35]. These include nano-lenses[36] and nano-antennas[37]. The localized surface plasmons in spherical core-shell structures can result in so-called spasers, with resonantly coupled transitions with excitons from dye molecules in the shell layer[38]. The coupling between localized surface plasmons from closely separated small particles results in electromagnetic hot spots[39-41]. Huge light scattering is found on anisotropic spherical particles with an active mechanism[42]. These remarkable results based on light scattering by small particles suggest many potential applications, including solar energy technologies[14, 43] and nano-scale lasers[44].

**Light Scattering by Spherical Particles in the Quasi-static Approximation and Beyond**

In the quasi-static approximation, light scattering by sphere particles is addressed by solving the Laplace equation for the scalar electric potential

$$\nabla^2 \Phi = 0 \text{ and } E = -\nabla \Phi, \tag{1}$$

with continuous boundary conditions

$$\Phi_1\big|_a = \Phi_2\big|_a \text{ and } \varepsilon_p \frac{\partial \Phi_1}{\partial r}\bigg|_a = \varepsilon_m \frac{\partial \Phi_2}{\partial r}\bigg|_a, \tag{2}$$

where $E = E_0 \bar{z}$, $a$, $\varepsilon_p$ and $\varepsilon_m$ are the electric field along $z$ direction, radius of particle, dielectric function of the particle and that of the medium, respectively. By comparing the scattering potential from the Laplace equation with that of a dipole, the effective dipole moment can be expressed as $P = 4\pi\varepsilon_m a^3 \frac{\varepsilon_p - \varepsilon_m}{\varepsilon_p + 2\varepsilon_m} E_0$. The cross-sections for scattering ($C_{sc}$) and adsorption ($C_{abs}$) are obtained from the scattering field radiated by this dipole, which is induced by the incident plane wave (i.e. the dipole moment $P$). The resulting expressions are[1]

$$C_{sc} = \sigma_{geom} Q_{sc}, \quad Q_{sc} = \tfrac{8}{3} q^4 \left| \tfrac{\varepsilon_d - 1}{\varepsilon_d + 2} \right|^2 \quad \text{and}$$

$$C_{abs} = \sigma_{geom} Q_{abs}, \quad Q_{abs} = 4q \, \text{Im}\left[ \tfrac{\varepsilon_d - 1}{\varepsilon_d + 2} \right], \tag{3}$$

where $\sigma_{geom} = \pi a^2$ is the geometrical cross-section. We introduce the dimensionless cross-sections of scattering ($Q_{sc}$) and adsorption ($Q_{abs}$), the dimensionless size $q = ka$ and the relative dielectric function $\varepsilon_d = \varepsilon_p / \varepsilon_m$ for convenience in the discussion that follows. In the formula, $k$ is the wave vector in medium.

The above formula implies that as the particle size is decreased, the efficiency of adsorption will dominate over the scattering efficiency. Therefore, particles with very small size will be difficult to detect by light scattering. In addition, one may note that there is a resonant enhancement for scattering and adsorption when the condition $\text{Re}(\varepsilon_d) = -2$ (the Fröhlich condition) is satisfied. This resonance is due to resonant excitation of the dipole surface plasmon. With the Drude model of dielectric function, the frequency of the dipole surface plasmon can be expressed as $\omega_{sp} \approx \omega_p / \sqrt{3}$, if the frequency of electron collisions $\gamma$ is small. The spectrum for the scattering cross-section then has a Lorentz line shape[27],

$$Q_{sc}^{(Ra)} = \frac{8}{3} \frac{\omega_{sp}^4}{(\omega^2 - \omega_{sp}^2)^2 + \omega^2 \gamma^2} q^4 \quad . \tag{4}$$

The above theory, however, doesn't capture the size effects on the spectra, including the changes of position and width of plasmon peak. Retardation effects in larger particles result in breakdown of the quasi-static approximation.

Within the Mie solution, the dimensionless cross-sections for scattering, extinction and absorption can be expressed as

$$Q_{sc}^{Mie} = \frac{2}{q^2} \sum_{l=1}^{\infty} (2l+1) \left\{ |a_l|^2 + |b_l|^2 \right\} \quad ,$$

$$Q_{ext}^{Mie} = \frac{2}{q^2} \sum_{l=1}^{\infty} (2l+1) \mathrm{Re}(a_l + b_l) \quad \text{and} \quad Q_{abs}^{Mie} = Q_{ext}^{Mie} - Q_{sc}^{Mie}. \tag{5}$$

The scattering amplitudes $a_l$ and $b_l$ are defined as follows,

$$a_l = \frac{F_a^e(l)}{F_a^e(l) + i G_a^e(l)} \quad \text{and} \quad b_l = \frac{F_b^m(l)}{F_b^m(l) + i G_b^m(l)}, \tag{6}$$

where $F_a^e(l), G_a^e(l), F_b^m(l)$ and $G_b^m(l)$ are related to the Bessel and Neumann functions. These general expressions are rather cumbersome to present here and can be found in literature[1]. Expanding the Bessel and Neumann functions in power series for small $q$, the functions $F_a^e(l), G_a^e(l), F_b^m(l)$ and $G_b^m(l)$ can be expressed as[27],

$$F_a^e(l) \approx q^{2l+1} \frac{(l+1)}{[(2l+1)!!]^2} \tilde{n}^l (\tilde{n}^2 - 1),$$

$$G_a^e(l) \approx \tilde{n}^l \frac{l}{2l+1} \left[ \tilde{n}^2 + \frac{l+1}{l} - \frac{q^2}{2} (\tilde{n}^2 - 1)(\frac{\tilde{n}^2}{2l+3} + \frac{l+1}{l(2l-1)}) \right],$$

$$F_b^m(l) \approx -\frac{\tilde{n} q^2}{2l+1} F_a^e(l),$$

$$G_b^m(l) \approx -\tilde{n}^{l+1} \left[ 1 + \frac{1-\tilde{n}^2}{2(2l+1)} q^2 \right], \tag{7}$$

where $\tilde{n} = \sqrt{\varepsilon_d} = n_d + i\kappa_d$ is the relative complex refractive index with the relative real refractive index $n_d$ and the relative absorption index $\kappa_d$. Near the dipole scattering resonance, only the term $l=1$ in Mie formula needs to be considered. For small particles, the magnetic scattering amplitudes $b_l$ can be ignored. Then the scattering cross-section becomes $Q_{sc}^{Mie} \approx 6|a_1|^2/q$, where $a_1 = a_l(l=1)$ is determined by equation (6). At the Fröhlich resonance condition in the formula for $F_a^e(l)$ and $G_a^e(l)$, the scattering cross-section due to the dipole resonance can be expressed as[27],

$$Q_{sc}^{dip-Mie} \approx \frac{8}{3} \frac{\omega_{sp}^4}{(\omega^2 - \omega_{sp}^2)^2 + \frac{4}{9} q^6 \omega_{sp}^4} q^4. \tag{8}$$

One may note that the $\gamma$ parameter (corresponding to dissipative losses due to electron collisions) is ignored in the derivation. The spectrum is found to have a Lorentz profile with an effective parameter $\gamma_{eff}$, which has a similar role as the dissipation parameter in the Rayleigh spectrum and can be expressed as $\gamma_{eff} = \frac{2}{3}\omega_{sp}q^3$. This damping is due to the radiative losses of plasmons. This is different from the dissipative loss term ($\gamma$) in Rayleigh formula. Therefore, the singularity in the corrected scattering formula is removed due to radiative losses, even if dissipative losses are neglected.

Intuitively, the red shift of the peak in the dipole resonance with increasing size is due to the weakening of the restoring force. This is because the distance between charges on opposite sides of the particle increases with size and so their interaction decreases. The red shift of resonance can be addressed directly by numerical solution of the Mie theory equations. The scattering cross-section, $C_{sc}$ from the dipole term neglecting dissipative losses is plotted in Figure 1a. The red shift is clearly seen. One may also note that the resonant scattering cross-section is similar for the different sizes. However, with dissipative losses, the scattering cross-section falls quickly with decreasing particle size (Figure 1b). Clearly, the effect of dissipative losses on the scattering increases with decreasing size. The effect of the dielectric function of the medium, $\varepsilon_m$ on the plasmon resonance of small metal particles is contained in the parameters $q$ and $\varepsilon_d$. From equation (3), $C_{abs}$ is proportional to $\frac{\varepsilon_m^{3/2} \text{Im}(\varepsilon_p)}{[\text{Re}(\varepsilon_p)+2\varepsilon_m]^2 + [\text{Im}(\varepsilon_p)]^2}$. The Fröhlich resonance condition then requires a decrease in Re($\varepsilon_p$), for an increase in $\varepsilon_m$. For most metals, such as Au and Ag, Re($\varepsilon_p$) decreases and Im($\varepsilon_p$) increases with decreasing frequency around the resonance. Therefore, the peaks in the absorption spectra are shifted to longer wavelengths and become broader and more intense with the increase in $\varepsilon_m$[22].

**Optical Scattering by Dielectric Particles**

The light scattering by dielectric particles is easy to calculate using the Mie theory. Figure 2a and 2b shows two typical scattering curves for particles of refractive index $n = 1.33$ and $n = 1.97$ (that of water and glass). The curves have a series of maxima and minima. Other dielectric spheres with different refractive indices show similar behavior. Finally, we note that in the limit of an extremely large particle the scattering cross-section is twice as large as the geometrical cross-section[19]. All these results for the scattering and extinction cross-sections were given fifty years ago[19]. These results, taken in the far field, are connected with the study of light transmission in mist, fogs, cloud chambers, and so on, but the near field regions for dielectric spheres have received attention much more recently.

For small size dielectric particles, the scattering cross-section will increase with increasing refractive index. This result follows from the Rayleigh theory and is indicative of the important role of dipole scattering. The squares of the intensity of the electric fields in the near field for different refractive indices are shown in Figure 3a-3c. The electric field intensity increases in the near field region with increasing refractive index. However, the near field electric field intensity from the dipolar mode doesn't become stronger as the size of dielectric particle decreases. Instead, the maximal value of electric field intensity increases along with a change in the spatial configuration, when the particle size increases in Figure 3d-3f.

Usually, Mie scattering will dominate when the particle size is larger than a wavelength. As shown in Figure 3d-3f, Mie scattering produces complex field distribution patterns reminiscent of directional antennas. This effect is sometimes referred to as a photonic nanojet. From the point of view of geometrical optics, the increase in the scattering field intensity with a more intense forward lobe reflects the fact that a dielectric sphere with large size behaves as a convex lens.

Localized regions of high intensity, such as those near the particle, can trap dielectric particles due to gradient forces (these are forces that arise because a particle with a dielectric constant higher than medium will lower the energy if it moves to the location of the maximum electric field intensity). This effect was observed by Ashkin et al. more than 30 years ago[45, 46].

The related optical tweezers technique, in which light is used to manipulate small particles, has been applied in many different fields, especially in medicine, biology and biophysics, where biologically inert particles can be functionalized and then manipulated using light[47-49]. Specifically, small dielectric particles can be picked up and controlled by tightly focused visible light lasers. It is interesting that this qualitative effect can also be understood just with ray optics[50].

A photon with energy $\hbar\omega$ carries the momentum $\hbar k$. If the photon is adsorbed by an object, a force $F$ on the object due to the transfer of momentum will be produced and given by the formula $F=n_m P/c$, where $n_m$ is the refractive index of the surrounding medium, $P$ is the power of light beam and $c$ is the light speed in vacuum. A dimensionless quantity $Q_{et}$ which is used to describe the efficiency of trapping light by a particular object with any shape, is defined by the formula $Q_{et} = n_m P/cF$. The refraction of light by a transparent object will result in the reaction force acting on the object, since the momenta of the photons are changed. For the plane perpendicular to the direction of light beam, an intensity profile with high symmetry will result in a force that will tend to move the object into the center of beam, since the force from the refraction, which points to the center, will provide a restoring force when the position of the object deviates from the center, shown in Figure 4. If the focus of the beam just is above the object, a force will be generated to lift the object up towards the focus.

The concepts of geometrical optics are simple and intuitively appealing, but they are not strictly applicable to the case of particles that have sizes below the wavelength. When one considers the electric field in the near field region, one sees that the force can be separated into two parts (note that both force terms necessarily imply a rate of momentum change of the light field). One is the force due to the intensity gradient and another one is from the scattering of light. The scattering force from the light beam with the intensity $I_0$ can be expressed as $F_{sc} = I_0 C_{sc} n_m / c$. The gradient force due to the intensity gradient can be given by the formula[46] $F_{gra} = -n_m^3 a^3 (\frac{\varepsilon_d - 1}{\varepsilon_d + 2}) \nabla (E^2) / 2$, where $E$ is the electric field near the particle. Clearly, the force is directed towards the higher intensity region and does not depend on the light propagation direction. The control of particles by optical trapping can be established when the gradient force exceeds the scattering force. In practice this is readily done for small particles, and precise three dimensional control of the particle position is possible by using optical interference or crossed beams to define a localized maximum in intensity.

An optically trapped particle in a viscous medium (viscosity $\eta$) will behave like a damped oscillator and thus the equation of motion will be[50], $m\ddot{x} + \beta \dot{x} + k_{st} x = 0$, where $k_{st}$ is the stiffness of optical trap and $\beta$ is the Stokes drag constant with the formula[51], $\beta = 6\pi a \eta$. Further expressions for dielectric spheres including scattering can be obtained using the explicit partial-wave representations[52].

This type technique based on gradient forces has been broadly applied in very diverse areas[53]. These range from disease diagnosis to gravitational detection. With new developments[54], optical tweezers can be used to detect biological compounds at the single-molecule level[55]. Control of the motion of particles at nanometer scales with piconewton forces enables studies of molecular and nanoscale dynamics, for example the investigations of molecular motors[56]. Furthermore, as mentioned and as shown in Figure 4, nanojets (directionally concentrated electromagnetic radiation) can be formed under large dielectric spheres. These nanojets are technologically important as they can be used to enhance Raman signals. Thus a dielectric particle, which is analogous to a nano-size convex lens, can induce a high light intensity under the particle (the nanojet), and this particle and its nanojet can be controllably moved and used as a nanoscale Raman probe using optical tweezers[57] (Figure 4f).

**Localized Plasmons of Metal Particles**

It may be noted in the above discussion that the scattering efficiency from the dipole term in Mie theory increases as the particle size decreases in the small size region around the resonance frequency (see the formula 8). This is clearly different from ordinary Rayleigh scattering. Ignoring dissipation, the high-order plasmon modes have resonant frequencies, $\omega_l^2 = \omega_p^2 l / (2l+1)$. Since all the amplitudes, $a_l$ tends to go to unity for the corresponding frequency, the scattering cross-sections of high-order plasmon modes can be expressed as $Q_{sc}(l) = \frac{2(2l+1)}{q(l)^2} = \frac{2(2l+1)^2}{l} \frac{c^2}{\omega_p^2 a^2}$. Since the resonance frequencies of different modes are different and the resonant peaks of different modes are limited, the total scattering cross-section for each resonant frequency is given by $Q_{sc} \approx Q_{sc}(l)$. Therefore, anomalous scattering of light with an inverse hierarchy of the resonances will occur if the dissipation term in the dielectric function is very small, as shown in Figure 5. Usually, with the condition of the actual dissipation[58], $\text{Im}\,\varepsilon_d(\omega_l) \ll \frac{q^{2l+1}}{l[(2l-1)!!]^2}$, the anomalous scattering rises. Note, however, that ordinary Rayleigh scattering is restored when the size parameter $q$ tends to zero.

The Fano resonance found in 1961[29] is well-known in quantum physics. Fano spectra arise from the constructive and destructive interference between a narrow resonant mode and a broad background spectral line. Fano spectra exhibit an asymmetric shape, specifically taking the form[25], $I(\omega) = \frac{(F\gamma + \omega - \omega_0)^2}{(\omega - \omega_0)^2 + \gamma^2}$, where $F$, $\omega_0$ and $\gamma$ are the Fano parameters, the position and the width of resonance, respectively. Fano resonances have been found in diverse quantum systems, such as quantum dots and tunnel junctions. Fano resonances are also expected to appear in light scattering.

In plasmonic materials, the resonant peak of each plasmon mode has a very different linewidth. Therefore, different plasmon modes can coexist in the same frequency region. Then Fano resonances can arise due to the constructive and destructive interference of plasmon modes with different multipolarity[30]. The resonant interference does not occur in the total optical cross-section, such as the scattering and extinction cross-section for a single particle. It is seen in differential scattering cross-sections, such as forward scattering (fs) and radar back scattering (rbs), with the formulas[25]

$$Q_{fs}^{Mie} = \frac{1}{q^2} \left| \sum_{l=1}^{\infty} (2l+1)[a_l + b_l] \right|^2, \text{ and}$$

$$Q_{rbs}^{Mie} = \frac{1}{q^2} \left| \sum_{l=1}^{\infty} (2l+1)(-1)^l [a_l - b_l] \right|^2. \quad (9)$$

The line width decreases quickly with increasing the order of plasmon mode, according to the formula[23], $\gamma_l = \frac{q^{2l+1}(l+1)}{[l(2l-1)!!]^2 (d\varepsilon_d/d\omega)_l}$. Clearly, the interaction of a dipolar mode and a quadrupole mode is easiest due to the radiative coupling, especially for the relatively small particles. The magnetic amplitude can be ignored, and then the low-energy interference of the electric dipole and quadrupole is given by formulas[25], $Q_{fs} = \frac{1}{q^2}|3a_1 + 5a_2|^2$ and $Q_{rbs} = \frac{1}{q^2}|3a_1 - 5a_2|^2$. $Q_{rbs}$ and $Q_{fs}$ as the functions of frequency are shown in Figure 6a, where a Fano resonance near the quadrupole resonance frequency is clearly seen.

The interference of incident and re-emitted light in the scattering process generates complex patterns in the near-field region. The energy flow, as represented by the Poynting vector, from the dipole has helicoidally shaped vortices, while that from the quadrupole is still more complex with vortices and singular points[27] (Figure 6b). Higher order modes can also interfere with the broad dipole mode as the size increases. However, it is important to note that the dissipative losses of plasmonic materials must be weak for the Fano resonance to appear, since the higher-order modes are rapidly suppressed when dissipative losses increase.

The Fano resonance of a single spherical particle is generally difficult to observe due to dissipative losses. If the widths and energy positions of plasmon modes can be modulated independently, the condition about the interference between a narrow discrete mode and a broad background resonance is easier to realize. An example is a nonconcentric ring/disk cavity[59, 60]. The dipolar modes from disk and ring interact to result in a hybridized bonding mode and a broad higher-energy anti-bonding mode[61]. The coupling between the quadrupolar mode from the ring and the anti-bonding dipolar mode due to the symmetry breaking of the nonconcentric geometry can induce an enhanced Fano resonance. Related ideas can be also applied to the other plasmonic nanostructures, such as nanoshells[62, 63], dolmen-type structures[64, 65], finite clusters of plasmonic nanoparticles[66-68], and so on. In addition, Fano resonances frequently appear in photonic crystals [69-71], such as periodic metallic structures on a single-mode slab. The waveguide mode from the slab can couple with the plasmon modes of the metallic structures exited by the incident light. Optical Fano resonances have recently been found in electromagnetic metamaterials[72-75]. The high asymmetrical profiles of Fano resonances suggests important applications, including novel sensors, as well as lasing and switching schemes[25].

In the case of non-magnetic particles, there are also other unconventional Fano resonances. An example is in the light scattering by small particles with large dielectric permittivity or with spatial dispersion[76]. This kind of resonance in scattering by small particles is beyond the applicability of Rayleigh approximation. The electromagnetic modes excited by the incident wave which can interfere with each other have the same multipole moment $l$. This results in the conventional Fano resonances, while those modes with different $l$, which have spatial dispersion, can yield directional Fano resonances[77].

Fano resonances can also occur in light scattering by magnetic particles. This occurs with negative magnetic permeability ($\mu < 0$) and positive dielectric permittivity ($\varepsilon > 0$). In that case, the interference of different magnetic multipole modes can result in the Fano effect, such as that between magnetic dipole ($b_1$) and quadrupole ($b_2$)[25]. With the effective magnetic permeability, the effect of magnetic modes on the light scattering becomes important and the interference of electric and magnetic modes (Kerker effect) can occur[78]. With the condition $\varepsilon=\mu$, the backward scattering gain is zero. It is also possible that the forward intensity is zero and the dipoles are out-of-phase, under the second Kerker condition. Excitingly, an unconventional forward–backward scattering asymmetry was recently observed experimentally in scattering by a single subwave-length sphere[26].

The Plasma is an important concept in physics. It is used to explain the energy losses of fast electrons in thin metal films. The theoretical work of Ritchie (1957)[79] and the experimental work of Powell and Swan (1959)[80] laid the groundwork for the study of surface plasmons by measurements of electron energy loss spectra. The optical properties of metallic materials in a low energy region are controlled mostly by collective plasmonic excitations of conduction electrons.

Surface plasmons can be excited by optical beams using a prism with the attenuated total reflection method as shown by Otto[81] and Kretschmann et al. in 1968 [82]. Importantly, in the case of metallic particles, the finite surface area can localize the propagation of light and result in localized surface plasmons, which as mentioned have many current and potential applications.

The dielectric function of an ideal bulk metal at low energy can be phenomenologically expressed by the Drude model of free electrons. In modeling real metals, a term corresponding to Lorentz oscillators is usually introduced to describe the increase of the imaginary part of dielectric function $Im(\varepsilon)$ due to the inter-band transitions[1]. These can also be calculated in detail using first principles electronic structure methods. According to Fermi liquid theory, the conduction bands of metals are continuous near the Fermi surface and the low energy properties are like those of an electron gas, though renormalized from the free electron gas and with anisotropy and other complexities reflecting the crystal lattice and band formation. As such, given the band structure, the plasma frequency can be directly calculated from the band dispersion at the Fermi surface and important insights about the nature of the metallic state can be gained from its comparison with experiment[83]. In non-cubic solids, the Drude plasma frequency has the form of a rank-2 tensor and therefore can be anisotropic. In any case, the presence of conduction electrons will result in intra-band excitations within the conduction band by the creation of electron-hole pairs.

For noble metals, such as gold and silver, there are also inter-band transitions from lower-lying d-bands to the sp-hybridized conduction bands. These are the main causes of dissipative losses. In addition, there are other generally weaker processes including elastic and inelastic electron scattering, such as electron-electron, electron-phonon and electron-defect interactions[84]. All these dissipative loss mechanisms can result in the non-radiative decay of plasmons and importantly can be phenomenologically described using the Lorentz-Drude dielectric model.

The dielectric functions of bulk gold and silver are shown in Figure 7 e and g[85]. Figure. 7 f and 7h show the spatial configurations of the square of electric fields of gold and silver particles with radius $R$=1.6 nm at the dipole resonance $Re(\varepsilon_d)$=-2. Clearly, the dissipative losses have an important effect on the intensity in the near-field. The variation of the near-field electromagnetic intensity configurations with particle size is shown away from resonance in Figure 7 a-7d. One may note that the strength of electric field in the near-field does not increase with this parameter and that this is obviously different from the behavior illustrated for dielectric particles in Figure 3d-3f.

The localized surface plasmon resonances of noble metal particles with the sizes of more than 10 nm have been well characterized experimentally[86]. The understanding of plasmon resonances for smaller sizes is, however, still poor. This is because both experiment and theory are challenging for small particle sizes[87, 88]. In particular, both quantum effects and detailed surface interactions become important as the electrons interact more strongly with the surface including the spill-over of conduction electrons at the cluster surface, which complicates geometrical analysis[89]. Quantitative predictions then require detailed calculations of the electronic structure for the actual atomic arrangements of the clusters of interest. For experiment, optical detection in the far-field becomes difficult for small particles due to the size dependent reduction in scattering intensity[1]. Theoretically, time dependent density functional theory (TDDFT) based methods[35, 90-92] are usually limited at present to particles with the sizes below 1-2 nm[93] but still useful insights have emerged. Methods that bring detailed quantum mechanical calculations to the longer length scales of interest would be very valuable in better understanding the size regime where quantum effects start to become important.

The first effect we mention is the red shift effect in the case of alkali-metal particles, which is due to the finite surface area[94, 95]. The red shift is understood in terms of the spill-over effect[96]. At small size, the electronic density profile will extend beyond the nominal surface. This is an effect of the high kinetic energy of the *s*-electrons that make up the conduction states of alkali metals. The resulting charge located outside the surface cannot be efficiently screened by the other electrons. So the polarizability is enhanced, which results in a decrease in the resonant frequency.

The effect of electron scattering at the surface may be described via a corrected dissipative loss term in the Drude model with the formula[89] $\gamma' = \gamma_{bulk} + \frac{A v_F}{R}$, where $\gamma_{bulk}$ is the parameter describing bulk dissipative losses, $R$ is the particle radius and $v_F$ is the Fermi velocity. $A$ is an empirical constant that can be set using fits of experimental data. This effect also results in a slight red shift of the resonant frequency.

Next we discuss the blue shift of the plasmon resonance of small non-alkali metal particles. This can be understood in terms $d$-electron contribution to the dielectric properties[84]. In bulk materials, the Lorentz term in Lorentz-Drude model represents the contribution of inter-band transitions, involving the $s$-$d$ interactions. The bulk plasmon resonant frequency is reduced from the unscreened value due to screening from $s$-$d$ interactions. For example, the bare plasmon energy of Ag is reduced from 9.2 eV to 3.76 eV by screening[96]. This is reduced at the surface of small particles as the $s$ electrons spill out. The reduced screening that results will then yield a blue shift (the surface-to-bulk ratio increases as the size decreases). Detailed quantitative characterization of spill-over effects on the surface plasmons of small particles will depend on the development of atomistic methods for the surface electronic structure and excitations that can be applied for cluster sizes of interest. It may be that interesting new effects will emerge from studies that include detailed surface structures and interactions.

Both top-down and bottom-up methods have been adopted to analyze the size-dependent plasmon frequency[97]. Starting with bottom-up approaches, cluster science has made an important contributions to the understanding of the optical properties of small particles both theoretically and experimentally[92, 96, 98-100]. From the top-down, plasmon resonances can be studied by aberration-corrected transmission electron microscope (TEM) imaging and monochromated scanning TEM electron energy-loss spectroscopy[101]. Microscopically, the free electron part of the Drude model can be modified to a phenomenological model of very small particles by considering the conduction electrons as an electron gas constrained in an infinite potential barriers[102, 103]. Then, the quantum size effects lead to a discrete set of energy levels near Fermi surface instead of a Fermi liquid. As discussed by Scholl et al., these quantum size effects result in the blue shift of resonance frequency[101]. This is in addition to the spill-out effect and the resulting weakened screening of $d$ electrons, as discussed above. It is, however, important to note that there remain inconsistencies between experimental results from top-down and bottom-up approaches. Methods that can span the full size range of interest will be very helpful in developing a more quantitative understanding of the size dependence[97].

This is an exciting time for nano-photonic applications based on light scattering by particles. For applications, tuning of properties is important. One avenue for this is through the use of core-shell particles, including the special case of hollow core particles, instead of simple single component particles. For the spherical case, one has two dielectric functions, the core radius and the particle (core+shell) radius as parameters, instead of the single dielectric function and radius as tuning parameters for the single component case. One example of a core-shell particle used in light scattering is the case of metal particles in an aqueous solution. In this case there may be chemical effects at the surface. In particular, the interface between particle and aqueous solution can be viewed as a double layer, and furthermore anodic or cathodic polarization can induce chemical changes due to anion adsorption or desorption, alloy formation, and metal deposition including deposition of a shell with a different composition (e.g. Ag on Pd)[104]. Light scattering in such cases can be dealt with using core-shell models. Core-shell particles can be used to obtain new optical properties that single spherical particles do not exhibit[105-108]. Furthermore, techniques for producing such particles are well developed[109, 110].

The core-shell model has been studied using the full solution of Mie theory[111] and also can be solved approximately using an electrostatic solution[112]. The surface plasmon resonance condition becomes Re($\varepsilon_{sh}\varepsilon_a+\varepsilon_m\varepsilon_b$)=0 with $\varepsilon_a= \varepsilon_{co}(3-2P_{ra})+2\varepsilon_{sh}P_{ra}$ and $\varepsilon_b= \varepsilon_{co}P_{ra}+\varepsilon_{sh}(3-P_{ra})$, where $\varepsilon_{co}$, $\varepsilon_{sh}$ and $\varepsilon_m$ are the dielectric functions of the core, the shell and the medium, respectively[112]. The parameter $P_{ra}$ is the ratio of the shell volume to the total volume of particle. The result is that the plasmon resonance frequency depends on the ratio of the core radius to the total radius of particle.

Core-shell structures also introduce the important concept of plasmon hybridization. This provides a powerful principle for the design of complex metallic nanostructures[113, 114]. The plasmon modes of nanoshells (core-shell particles with an empty core, i.e. hollow shells) can be viewed as arising from hybridization of the plasmon modes of a nanoscale sphere and a cavity[114]. This hybridization results in a low-energy bonding mode and high-energy antibonding mode, as mentioned in relation to the Fano effect. Many non-trivial nanostructures, such as gold nanostars[115] and nanorice[116] have plasmons that can be understood in terms the interaction of the coupled plasmons of simpler systems[117].

The interparticle distance is another variable that can be used to produce new physics and applications. Examples are quantum tunneling[118] and large electromagnetic enhancements at the junctions[119]. The development of nanoscale fabrication methods has made possible the production of different forms of nanoparticle arrays[66, 67, 118, 120]. These include dimers, chains, clusters, and uniform arrays. The simplest prototype, which can be used as a model, is a nanoparticle dimer. The interaction between localized plasmons and the interference of the electromagnetic fields from these plasmons are the two major factors that control the electromagnetic enhancements at the junctions. Different methods, such as the coupled dipole approximation (CDA)[120], the finite difference time domain (FDTD) method[121] and plasmon hybridization[122], have recently been used to understand the plasmonic properties of dimers. For the practical calculation, the temporal couple-mode model as an effective method is also developed[123, 124]. Within the framework of the hybridization concept, the dimer plasmons can be treated as bonding and antibonding combinations of the single particle plasmons. The shifts of the plasmons at large interparticle distance then follow the interaction between two classical dipoles, since this is the interaction that leads to the hybridization. At shorter distances, the plasmon shifts in dipolar models become stronger and vary more rapidly with distance. This is a consequence of hybridization (or mixing) coming from higher multipoles[122]. In addition, new interesting effects beyond the hybridization models, such as Young's interference, have been recently observed in the plasmonic structures[125].

The plasmon modes for the symmetric nanoclusters can be analyzed based on plasmon hybridization with group theory[66]. In addition, by introducing the symmetry breaking, nonsymmetrical nano clusters can also be analyzed. In the case of uniform two-dimensional nanoparticle arrays with coupling to localized plasmons can result in a coherent interaction of the array with light propagating in the plane of the array. This results in a plasmonic band structure[126-129]. In addition, in the subwavelength nanostructures, there is an substantial opportunity to obtain the superscattering if one can maximize the contributions from different channels[130]. These can be enabling for a number of applications, including various photonic metamaterial applications and plasmonic lasers[131, 132].

The polarization of a particle array can be expressed in the simple dipole approximation as $\alpha_{arr} = (1/\alpha - S)^{-1}$, where $\alpha$ and $S$ are the polarization of a single particle and the structure factor of array, respectively[133]. There will be a geometric resonance when the wavelength of scattering light is commensurate with the periodicity of the particles array[134]. The study on light scattering of uniform arrays of nanoparticles is strongly connected with the fields of photonic crystals and metamaterials. A detailed review was given by Garcia de Abajo[135], to which we refer the reader for details.

Finally, we note that nonlinear optical responses can be very strongly increased using nanoparticle plasmons. This is by two main mechanisms, namely through the field enhancement near the particle surface and via the sensitivity of resonance frequency to the dielectric function of surrounding medium[136]. Some of the first work on nonlinear-optical effects of small metallic particles was on nanoparticle colloids[137]. Extension of Maxwell-Garnett theory for the low-concentration limit of particles in the medium ($C_{ra} \ll 1$) can be used. The effective dielectric function of the nanoparticle colloids can be expressed as[22] $\varepsilon_{eff} = \varepsilon_m + 3C_{ra}\varepsilon_m \frac{\varepsilon_p - \varepsilon_m}{\varepsilon_p + 2\varepsilon_m}$. The resulting third-order susceptibility $\chi_m^{(3)}$ from plasmonic enhancement then can give rise to substantial optical Kerr effects[138, 139]. The formal electromagnetic description of small particle second-harmonic scattering (hyper-Raleigh, which should vanish in the dipole approximation due to inversion symmetry) was given by Dadap et al.[140], who described the second harmonic generation on a small centrosymmetric sphere on the basis of the Mie theory and determined the non-linear susceptibilities and radiation pattern. This formalism, although based on the local bulk response, provides an approach for dealing with the contributions from nonlocal dipole and other multipole modes. Giant second-harmonic scattering has been observed in experiments on suspensions of small gold particles[141], and even for individual gold nanoparticles[142]. Effective second-harmonic generation has also been studied in plasmonic structures with low symmetry, such as gold nanocones with a sharp tips[143], nano-apertures surrounded by gratings[144], and non-centrosymmetric gold nanocups[145].

**Spasers: Surface Plasmon Lasing from Active Particles**

In lasers, coherent electromagnetic energy is generally concentrated in the active region by a Fabry-Perot or similar resonator. Let us consider an ideal resonator with two perfect mirrors. The minimal distance between two mirrors for storing the electromagnetic field is half of the wavelength, as shown in Figure 8b. However, by using resonance between the electromagnetic field and a surface plasmon (described by a dipole model for a small metal particle), the electromagnetic energy can be concentrated in a much smaller region. The particle size for this can be estimated from the skin depth given by[44] $D_s = \frac{\lambda}{2\pi}[\mathrm{Re}(\frac{-\varepsilon_p^2}{\varepsilon_p + \varepsilon_m})^{1/2}]^{-1}$, where $\lambda$ is the vacuum wavelength. In the optical region, the size ($D_s$, corresponding to the so-called nano-plasmon) for noble metals, such as silver, gold and copper, is roughly 25 nm[44]. The minimum size of a nano-plasmonic system is determined by the distance electrons move on the surface in a period of the optical wave. This is given by the formula[146] $L_{nl} \sim v_F/\omega$, where $v_F$ is the effective Fermi velocity.

The resonance between a surface plasmon and the electromagnetic field also results in the loss of energy due to the dissipative losses (from the imaginary part of the dielectric function). This induces a loss of the plasmonic field with a decay rate $\gamma_p \propto (\mathrm{Im}\,\varepsilon_p)^{-1}$. For a metal, a quality factor is defined by $Q_p = \frac{-\mathrm{Re}[\varepsilon]}{\mathrm{Im}[\varepsilon]}$, where $\varepsilon$ is the dielectric function of the metal. This provides a criterion that can be used to estimate whether a given substance is a good plasmonic material in an optical region of interest. The quality factors of gold and silver are shown in Figure 8a. In addition, optical radiation also results in the loss of plasmonic field according to the formula[44], $\gamma_p^r = 4\varepsilon_m^{3/2}(\frac{\omega_{sp}a}{c})^3[\frac{\partial \mathrm{Re}\,\varepsilon_p(\omega_{sp})}{\partial \omega_{sp}}]^{-1}$.

Surface plasmons can be amplified by the stimulated emission of radiation in the presence of a gain medium, such as dielectric materials containing excited dye molecules. Lasing depends on the presence of two principal conditions − a cavity for the resonant generation of coherent optical modes and medium with gain due to population inversion. A realization, the so-called spaser (surface plasmon amplification by stimulated emission of radiation) consists of small plasmonic metal particles surrounded by a gain medium such that the linewidth of the light emission from the gain medium overlaps with that of plasmon mode[147]. Thus, the losses of energy of the plasmonic particles due to dispersion and radiation can be compensated by light emission of gain medium. The key point is that by using a composite of plasmonic metal particles in a gain producing medium one may obtain a coherent radiation field in a lasing system that is smaller than the wavelength of the light.

Stockman used a quantized treatment of a gain medium and a quasi-classical treatment of the surface plasmon to construct a semi-classical dynamical equation for the spaser[148]. This gives a description of the spaser mechanism as shown in Figure 8c and 8d. The equation of motion of the density matrix of a *p*-th chromophore (the active elements in the gain medium) can be expressed as $i\hbar\dot{\rho}^{(p)} = [\rho^{(p)}, \hat{H}]$, with $\hat{H} = \hat{H}_g + \sum_n \hbar\omega_n \hat{a}_n^+ \hat{a}_n + \sum_p \hat{E}(r_p)\hat{d}^{(p)}$, where $\hat{H}_g$ is the Hamiltonian of gain medium. $r_p$ and $\hat{d}^{(p)}$ are the coordinate vector and dipole moment operators of the p-th chromophore in gain medium, respectively. The second term in the total Hamiltonian is for the quantized surface plasmon and the third term is for the coupling between the chromophore and the electric field of surface plasmon. By defining the gain factor of chromophore, which is a two energy level system with the formula, $g = \frac{4\pi\omega\sqrt{\varepsilon_m}}{3c\hbar\Gamma_{12}}|d_{12}|^2 n_c$, where $d_{12}$ is the transitional dipole element between ground state and excited state, the threshold gain for spasing can be obtained and expressed as[44]

$$g_{th} = \frac{\omega}{c\sqrt{\varepsilon_m}} \frac{\text{Re}[s(\omega)]}{1 - \text{Re}[s(\omega)]} \text{Im}[\varepsilon_p(\omega)]. \tag{10}$$

Here $\Gamma_{12}$ is the rate constant for describing the polarization relaxation, $n_c$ is the density of the chromophores, and $s(\omega) = \frac{\varepsilon_m}{\varepsilon_m - \varepsilon_p}$ is Bergman's spectral parameter[149]. One may note that the threshold for gain just depends on the spasing frequency and the dielectric properties of system. However, the spasing frequency is determined by the geometry of the system.

Progress in spasers, including both theory and experiment, has been rapid[44]. The original theoretical concept of the spaser was proposed with V-shaped metallic structures and semiconductor quantum dots[10]. Following this, a nanolens spaser was proposed with a linear chain structure of metal nanospheres combined with an active medium[150]. A proposal for a spaser based on metal cores with an active shell was considered on basis of linear electrodynamics[151]. A narrow-diversion coherent radiation on based on the combination of a metamaterial and a spaser was proposed by Zheludev et al.[152]. The combination of the plasmon of an anisotropic spherical particle and an active medium was also proposed to result in a spaser[42]. In experiments, a spaser was realized on a conjugate structure based on a metallic core and a dye-doped dielectric shell[38] (Figure 8). Spasers have also been realized in other nano-structures, such as in CdS nanowires combined with a silver substrate and separated by a MgF$_2$ layer[153]. The development of active sub-wavelength optical elements such as in spasers, is expected to lead not only to diverse applications, but also to new fundamental insights into non-linear light matter interactions.

**Conclusions**

We have briefly reviewed the theory of light scattering by small spherical particles and aspects of the important progress on light scattering on small spherical particles. It is remarkable that although many of the fundamental aspects of the theory are more than 100 years old, there continue to be new, surprising and useful developments, such as spasers and optical tweezers based on it. The interest 100 years ago was in the far field. While the formalism showed fascinating near field behavior, specifically giant concentrations of electromagnetic energy in regions much smaller than the wavelength and with complex spatial distributions, this was not explored until much more recently. Now these effects are being exploited to yield remarkable new nanoscale effects and potential applications in different science areas, such as high-resolution optical imaging, small-scale sensing techniques, light-activated cancer treatments, enhanced light absorption in photovoltaics and photocatalysis, and numerous biomedical applications. We expect that many more applications will be developed exploiting optical scattering by small particles, and especially nanophotonics applications based on the near field, and far field applications using linear and non-linear plasmonic effects.

**Acknowledgements:**

Work at ORNL was supported by the United States Department of Energy, Basic Energy Sciences, Materials Sciences and Engineering Division

**Fig.1**

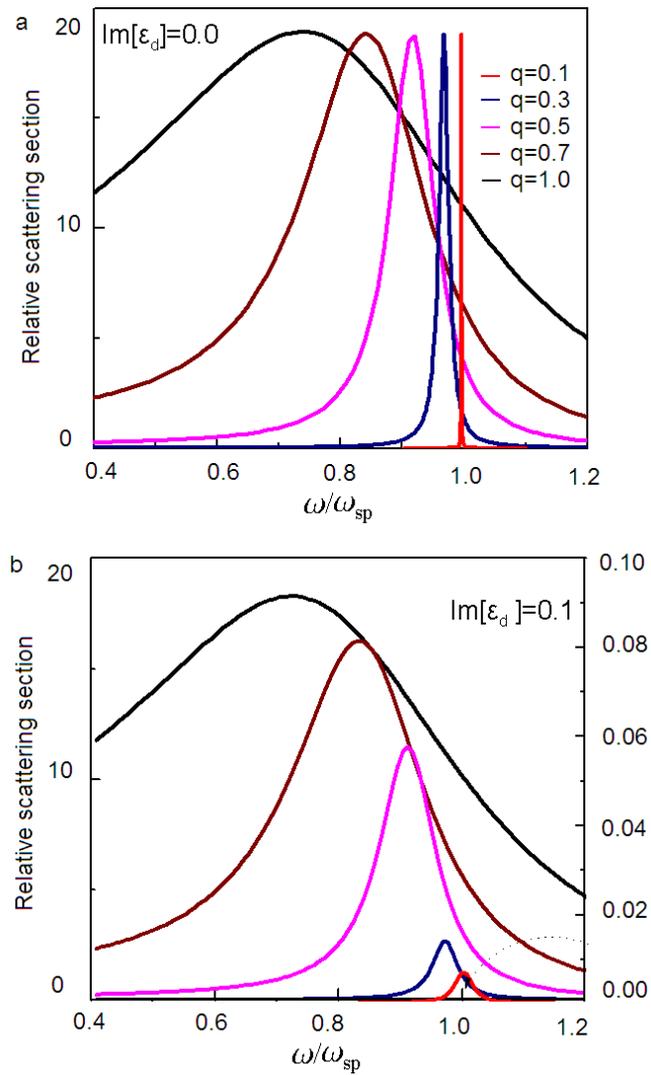

Figure 1 Illustration of the red shift of the dipole resonance, showing the relative scattering cross-section as a function of frequency for the different size parameters q in the cases of Im[$\varepsilon$]=0 (a) and Im[$\varepsilon$]=0.1 (b).

**Fig.2**

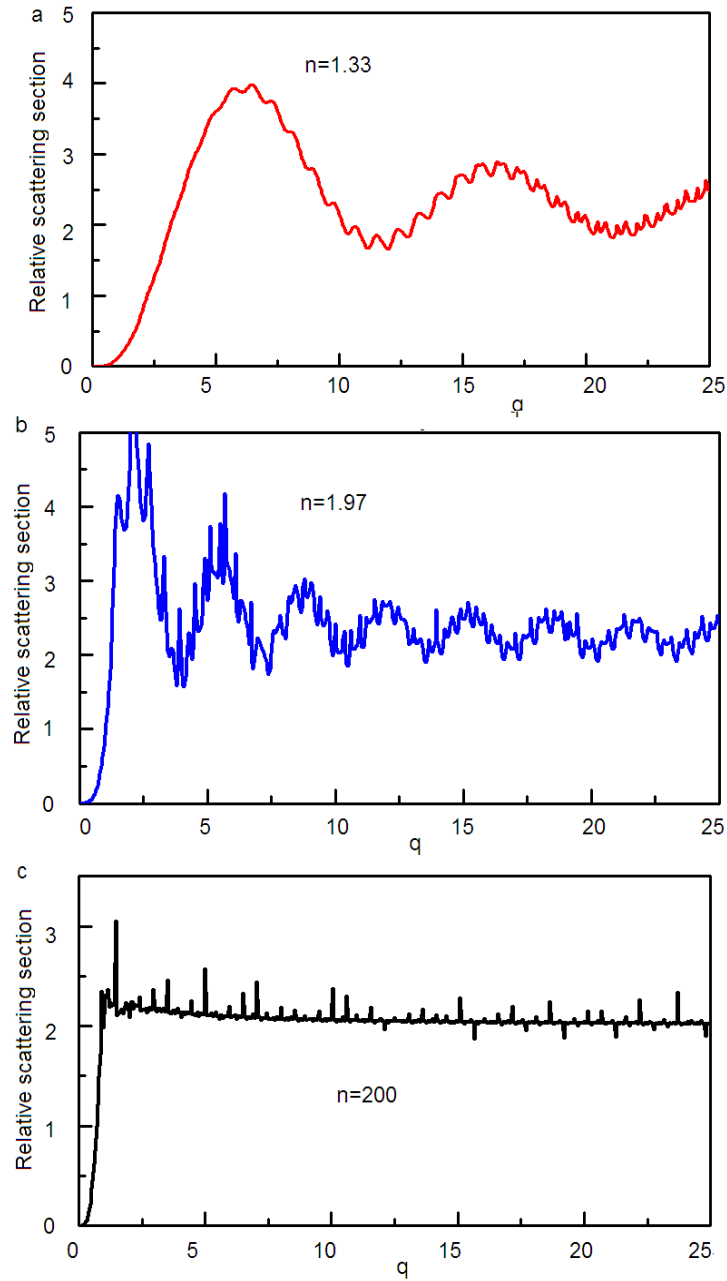

Figure 2 Relative scattering cross-sections of a water particle with reflective index $n$ =1.33(a), glass particle with reflective index $n$ =1.97 (b), and ideal particle with very high reflective index $n$ =200 (c) as the functions of dimensionless size parameter $q$. The first maximum of $Q_{sc}$ happens for the quantity $2q(n-1) \sim 4$. The case of refractive index $n$ =200 represents $n \sim \infty$. $Q_{sc}$ approaches $2\pi a^2$ at large $q$.

**Fig.3**

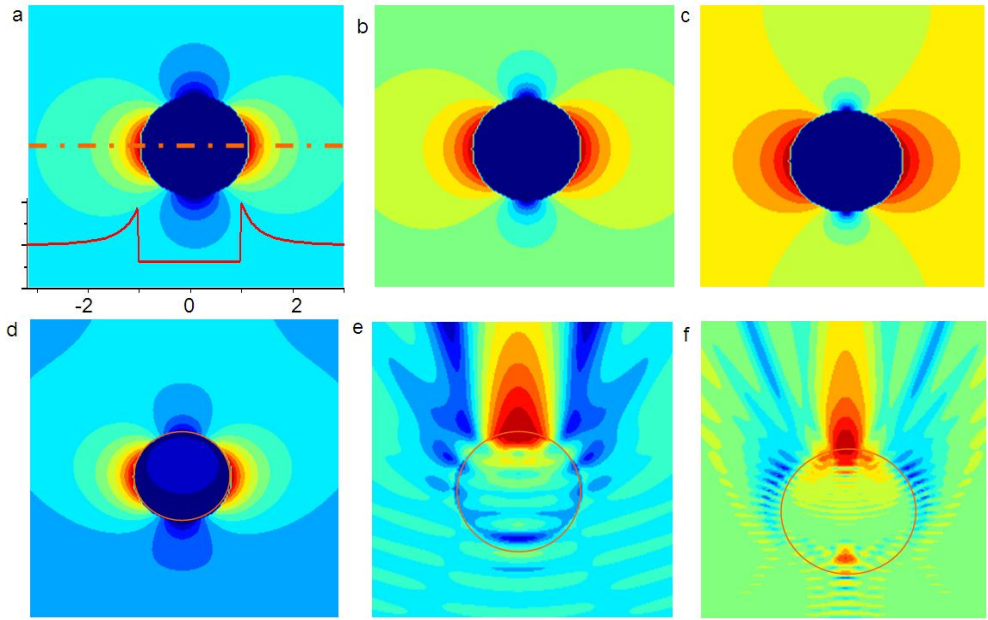

Figure 3 Near-field distribution of the square of electric field intensity for different dielectric particles including the small particles with the radius $R$=1.8 nm and different reflective indexes $n$ = 1.33 (a), 2.5 (b) and 4 (c) scattering light $\lambda$=496 nm and with the reflective indexes $n$ = 1.33 and different radii $R$=50nm (d), 500 nm (e) and 1000 nm (f) scattering light $\lambda$ = 500 nm. In (a-f), $E_{max}^2$ = $1.90E_0^2$, $5.15 E_0^2$, $7.10 E_0^2$, $2.20E_0^2$, $16.0 E_0^2$, $38.0E_0^2$, respectively.

**Fig. 4**

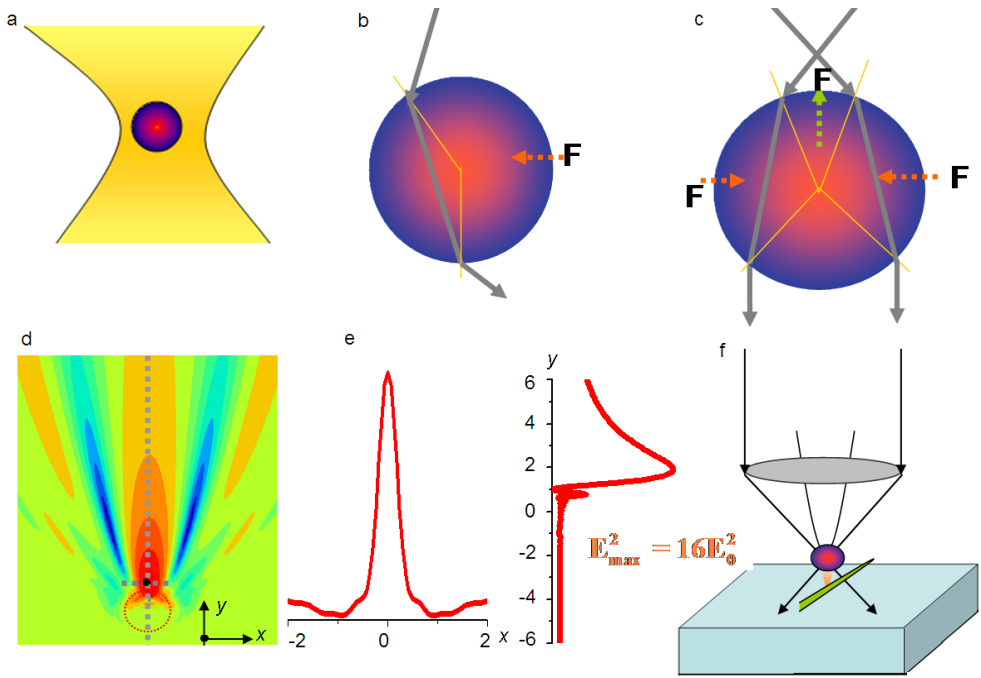

Figure 4 Schematic presentation of an optical tweezer (a), optical force from the view point of optical rays (b, c), Using the combination of optical trapping and nanojets for optically scanning samples on the surface of substrate (f), and the near-field distribution of the square of electric field density of a glass sphere ($n$ =1.5) in water ($n$ =1.33) with radius $R$=800 nm scattering light $\lambda$ = 500 nm (d) and the distribution of $|E|^2$ along the dotted lines in (d) with $E_{max}^2 = 16E_0^2$ (e).

**Fig. 5**

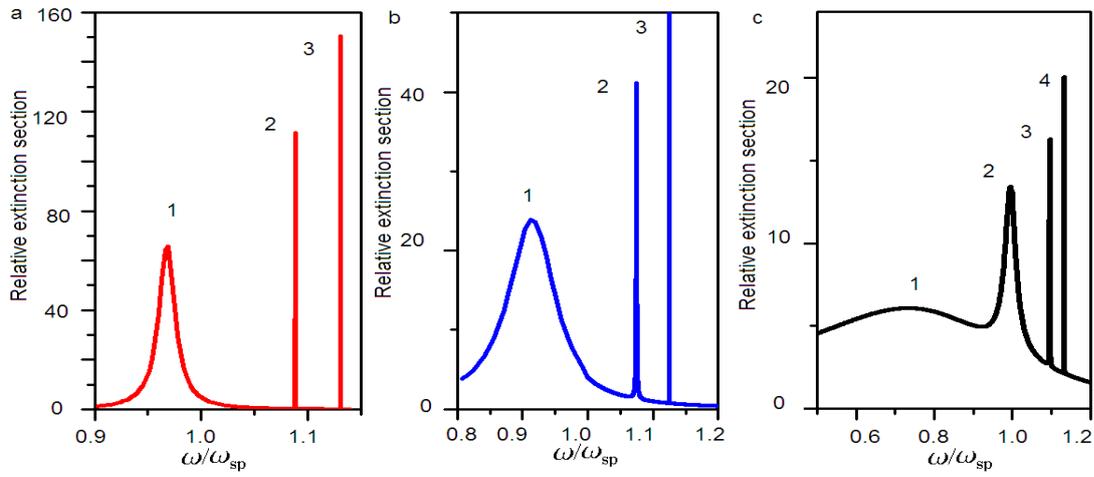

Figure 5 Relative extinction cross-section of a particle in the non-dissipative limit as a function of frequency for the different $q = 0.3$ (a), $q = 0.5$ (b) and $q = 1.0$ (c). Note that the dielectric function is with the model $\varepsilon(\omega)=1-3(\omega_{sp}/\omega)^2$, where $\omega_{sp}$ is the resonance frequency of the dipole model in the limit of small $q$.

**Fig. 6**

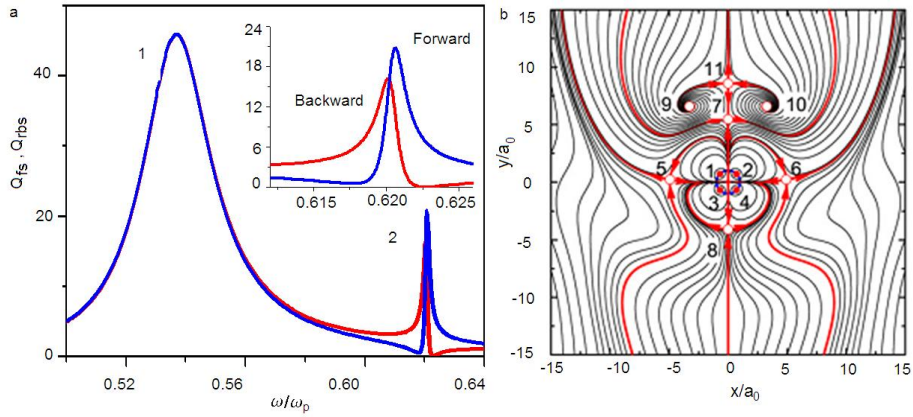

Figure 6 Fano resonance on Mie scattering by a small metal sphere (a) and energy flow in the quadrupole resonance with the singular vortices represented by the Poynting vector field (b). In (a), Radar back scattering and forward scattering directions are indicated by the red line and blue line, respectively, the dielectric function $\varepsilon(\omega)$ is described by Drude model with the dissipation parameter $\gamma=0.001\omega_p$ and the particle size is $a = 0.8c/\omega_p$. In (b) $q = 0.3$ and $\varepsilon_d = -1.553$, the blue line denotes the particle surface, the red lines indicate the separatrix. Figure reproduced with permission from ref. 27 : b, ©2007 IOP.

**Fig. 7**

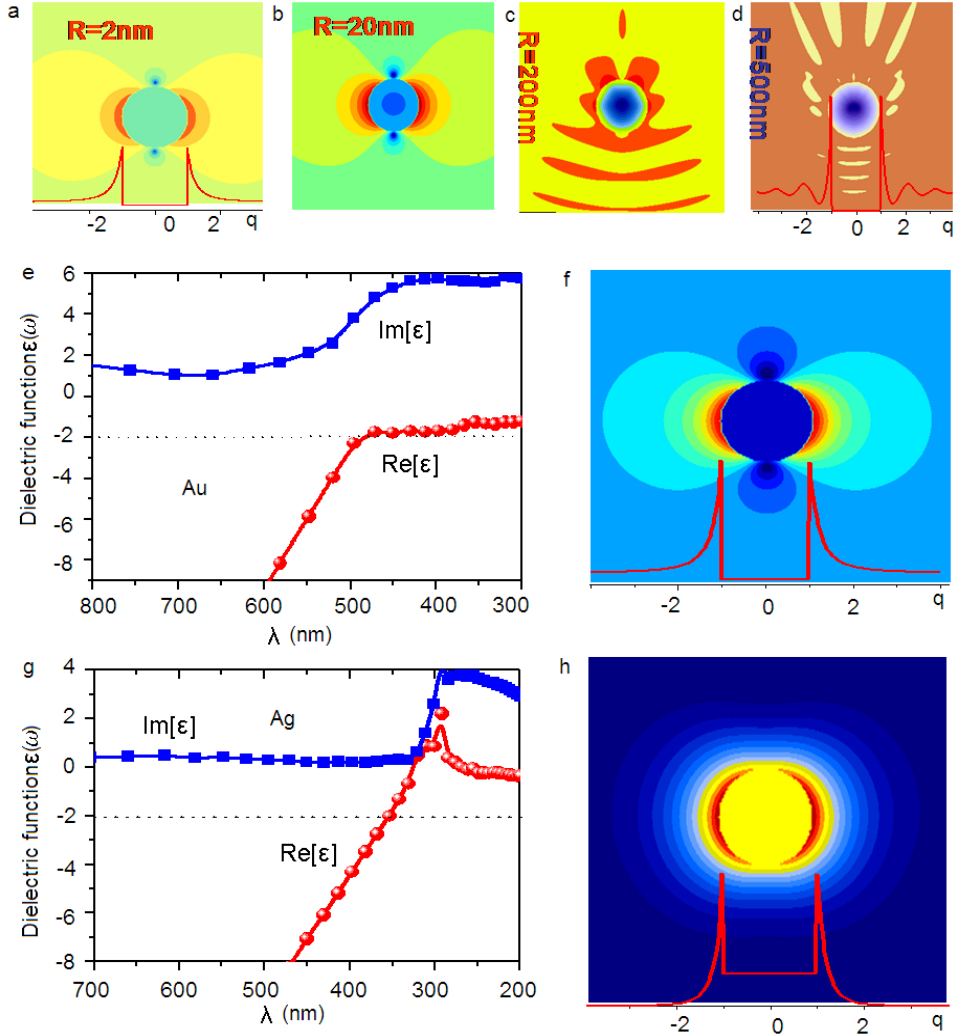

Figure 7 Near-field distribution of the square of electric field density on a silver particle scattering light $\lambda=496$ nm away from resonance with different sizes (a-d), where the dielectric constant is $\varepsilon= -9.56 + 0.31i$ (the radius R=2 nm, $E_{max}^2=14\ E_0^2$; R=20 nm, $E_{max}^2=16\ E_0^2$; R=200 nm, $E_{max}^2=17\ E_0^2$; R=500 nm, $E_{max}^2=16\ E_0^2$), dielectric function and near-field distribution of electric field density on a small particle at the resonance condition for gold (e, f) and silver (g, h). In (f), $\lambda=481$ nm, the radius of sphere is R=1.6 nm, dielectric constant is $\varepsilon=-2.0 +4.4i$ and $E_{max}^2=11\ E_0^2$. In (h), $\lambda=354$ nm, the radius of sphere is $R=1.6$ nm, the dielectric constant is $\varepsilon=-2.0 + 0.28i$ and $E_{max}^2=457\ E_0^2$.

Fig. 8

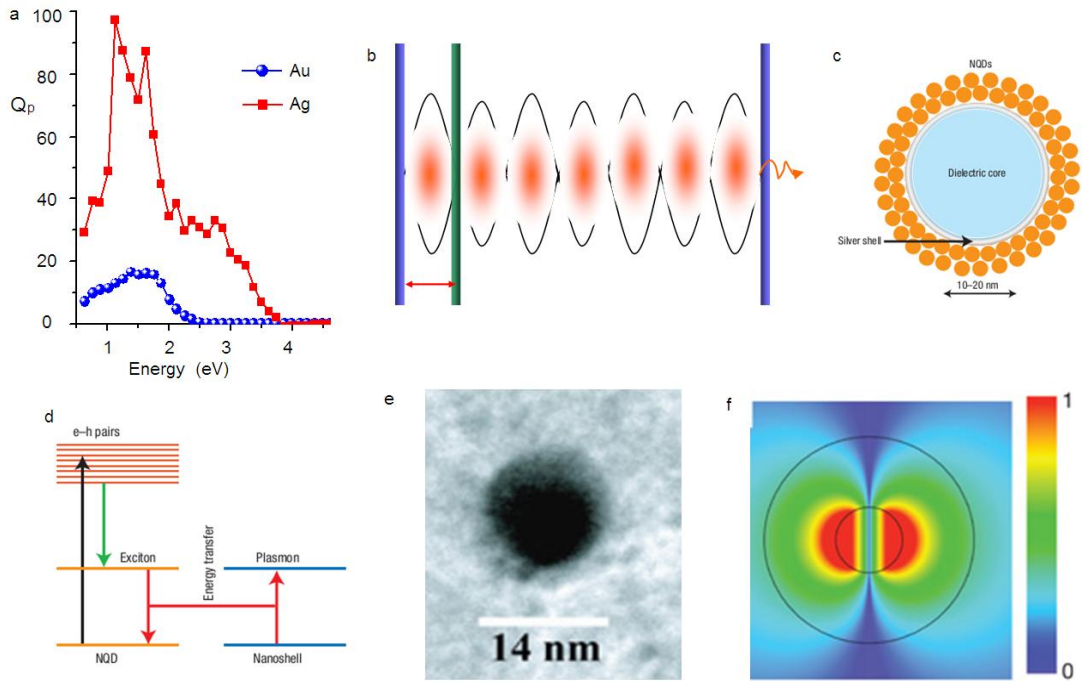

Figure 8 Quality factor of plasmons of gold and silver (a), schematic presentation of the electromagnetic wave in a Fabry-Perot resonator with the minimum possible distance between two mirrors (b), and the spasing mechanism (c-f). Schematic structure of a core-shell with a silver nanoshell on a dielectric core surrounded by the dense nanocrystal quantum dots (c), schematic presentation of the mechanism for energy transfers between e-h pairs and a resonant plasmon model via an exciton level (d), transmission electron microscope image of a Au core in a Au/silica/dye core–shell nanoparticle structure (e), and spaser mode with $\lambda = 525$ nm and $Q = 14.8$; the inner and outer circles represent the core and the shell layers, respectively. Figure reproduced with permission from ref.147 : c, d, ©2008 NPG; ref. 38: e, f, ©2009 NPG.